\documentclass[aps,prb,showpacs,twocolumn,superscriptaddress]{revtex4-1}
\usepackage{graphicx,amsmath,amssymb}
\usepackage[usenames]{color}
\usepackage{mathrsfs}
\usepackage{indentfirst}
\usepackage{float}
\usepackage{braket}
\usepackage[colorlinks=true, citecolor=blue, urlcolor=blue, linkcolor=blue ]{hyperref}
\hypersetup{breaklinks=true}
\begin{document}

\bibliographystyle{apsrev4-1}

\title{Ellipticity dependence transition induced by dynamical Bloch oscillations}
\author{Xiao Zhang}
\affiliation{Center for Interdisciplinary Studies and Key Laboratory for Magnetism and Magnetic Materials of the MoE, Lanzhou University, Lanzhou 730000, China}%

\author{Jinbin Li}
\affiliation{School of Nuclear Science and Technology, Lanzhou University, Lanzhou 730000, China}

\author{Zongsheng Zhou}
\affiliation{Center for Interdisciplinary Studies and Key Laboratory for Magnetism and Magnetic Materials of the MoE, Lanzhou University, Lanzhou 730000, China}

\author{Shengjun Yue}
\affiliation{School of Nuclear Science and Technology, Lanzhou University, Lanzhou 730000, China}

\author{Hongchuan Du}
\email{duhch@lzu.edu.cn}
\affiliation{School of Nuclear Science and Technology, Lanzhou University, Lanzhou 730000, China}

\author{Libin Fu}
\affiliation{Laboratory of Computational Physics, Institute of Applied Physics and Computational Mathematics, Beijing 100088, China}

\author{Hong-Gang Luo}
\email{luohg@lzu.edu.cn}
\affiliation{Center for Interdisciplinary Studies and Key Laboratory for Magnetism and Magnetic Materials of the MoE, Lanzhou University, Lanzhou 730000, China}
\affiliation{Beijing Computational Science Research Center, Beijing 100084, China}

\date{\today}

\begin{abstract}
The dependence of high-harmonic generation (HHG) on laser ellipticity is investigated using a modified ZnO model. In the driving of relatively weak field, we reproduce qualitatively the ellipticity dependence as observed in the HHG experiment of wurtzite ZnO. When increasing the field strength, the HHG shows an anomalous ellipticity dependence, similar to that observed experimentally in the single-crystal MgO. With the help of a semiclassical analysis, it is found that the key mechanism inducing the change of ellipticity dependence is the interplay between the dynamical Bloch oscillation and the anisotropic band structure. The dynamical Bloch oscillation contributes additional quantum paths, which are less sensitive to ellipticity. The anisotropic band-structure make the driving pulse with finite ellipticity be able to drive the pairs to the band positions with larger gap, which extends the harmonic cutoff. The combination of these two effects leads to the anomalous ellipticity dependence. The result reveals the importance of dynamical Bloch oscillations for the ellipticity dependence of HHG from bulk ZnO.
\end{abstract}                         

\maketitle

\section{Introduction}
The ellipticity dependence of high-harmonic generation is an important and fundamental issue in strong-field physics, which has been studied both theoretically and experimentally for gaseous media since 1990s \cite{Corkum2007,RevModPhys.81.163,PhysRevA.48.R3437,PhysRevA.50.R3585,PhysRevLett.80.484}. With increasing ellipticity \cite{PhysRevLett.74.2933,PhysRevA.51.R3418}, it was observed that the yields of gas harmonics descend rapidly, which confirmed the recollision mechanism of gas-HHG. Based on this mechanism, people creatively proposed the polarization gating and double optical gating to produce isolated attosecond pulses \cite{Corkum94,Sansone443,Lijie2017}. In recent years, HHG experiments have been extended to crystal materials \cite{Ghimire2010,LuuT2015,Vampaa2015}, which shows much rich and/or different ellipticity-dependent behaviors in comparison to the gas-HHG. For example, in rare-gas solids \cite{Georges2016} the harmonics exhibit an atomic-like ellipticity dependence, and in bulk ZnO \cite{Ghimire2010} the emitted harmonics are less sensitive to ellipticity. However, it is surprising that in single-crystal MgO \cite{You2016} the HHG shows an anisotropic and anomalous ellipticity dependence. Even for two-dimensional (2D) materials, the situation is also complicated. In monolayer MoS$_2$, the harmonic yields are suppressed monotonously with increasing ellipticity \cite{LiuH2016}, but for graphene it becomes enhanced by elliptically polarized light \cite{Yoshikawa736}. These intriguing experimental observations have attracted much theoretical attention \cite{PhysRevA.93.043806,PhysRevLett.116.016601,PhysRevB.94.241107,Nicolas2017,PhysRevA.97.063412,PhysRevA.96.063412}. However, how to get a clear physical picture to understand the different ellipticity-dependent behaviors remains open.

In contrast to a gaseous medium, the motion of an electron in a solid is affected strongly by the periodic structure of the crystal lattice. Once a constant electric field drives an electron to the boundary of the Brillouin zone (BZ), it will experience a Bragg reflection on the same band or Zener tunneling to a neighboring conduction band \cite{PhysRev.57.184,PhysRevA.70.052708,BZO2006,PhysRevLett.102.076802,SHEVCHENKO20101,RevModPhys.90.021002}. The Bragg reflection of the electron on a single band is known as the Bloch oscillation (BO) \cite{Bloch1929,Zener523,Foldi2013}. If the external field is time dependent, the similar phenomenon in strong-field physics is called the dynamical Bloch oscillation (DBO) \cite{Foldi2013,Schubert2014}. In the early studies of intense laser and solid interaction, the DBO was considered as one of the main mechanisms of generating solid harmonics \cite{Schubert2014,PhysRevLett.107.167407,PhysRevA.85.043836}. For multiple band systems, recent works \cite{PhysRevB.94.075307,PhysRevA.97.043413} show that a coherent superposition of dynamical Bloch oscillations and Zener tunneling, i.e., Bloch-Zener oscillation (BZO), has significant influences on the HHG. This is because the DBO changes the group velocity of carriers rapidly, which affects the recombination of electron-hole pairs \cite{Zaks2012}. We believe that the DBO would also play an important and essential role in ellipticity dependence of solid-HHG. This is our main motivation of the present work.

Focused on the influence of the DBO on ellipticity dependence of HHG, it is helpful to firstly make some arguments in a simple two-band system using the picture of electron-hole recollision \cite{vampa2016role}. In the absent of dynamical Bloch oscillations, the excited carrier experiences two velocity reversals during one optical cycle, and only the electron-hole pairs excited after the peak of the field can recombine. When the DBO exists, the Bragg reflection could make the pairs oscillate and collide multiple times in real space, which should generate the new quantum paths. It is natural to associate that the sensitivity of these new quantum paths to ellipticity may be quite different from the old ones, and those new paths could change the ellipticity dependence of harmonics. To further test our idea, we use a modified ZnO model and calculate explicitly the harmonic spectra at different laser ellipticies. In a relatively weak field strength, we find that the harmonics of different orders are gradually suppressed in the same way with increasing the ellipticity, which is qualitatively consistent with the experimental observations of ellipticity dependence in bulk ZnO \cite{Ghimire2010}. When increasing field strength, the HHG shows interesting features that the lower-order harmonics are suppressed rapidly and the higher-order ones are enhanced as increasing ellipticity. This is a typical anomalous ellipticity dependence, similar to that observed experimentally in the single-crystal MgO. Thus we uncover that the interplay between the DBOs and anisotropic band structures can lead to the transition of ellipticity dependence in solids and find a possible link between different ellipticity dependence in solids.

\section{Theoretical approach}
\subsection{Density matrix equations}
Our simulation of laser-solid interaction is based on density matrix equations \cite{PhysRevLett.113.073901,PhysRevB.91.064302,vampa2016role} (atomic units are used throughout this paper):
\begin{subequations}
	\begin{equation}
	\dot{n}_{m}=i\sum_{m'\ne m}\Omega_{mm'}\pi_{mm'}e^{iS_{mm'}}+c.c.,
	\end{equation}
	\begin{equation}
	\begin{split}
	\dot{\pi}_{mm'}=&-\frac{\pi_{mm'}}{T_{2}}+i\Omega_{mm'}^{*}(n_{m}-n_{m'})e^{-iS_{mm'}}
	\\&+i\sum_{m''\notin\{m,m'\}}(\Omega_{m'm''}\pi_{mm''}e^{iS_{m'm''}}
	\\&-\Omega_{mm''}^{*}\pi_{m'm''}^{*}e^{-iS_{mm''}}),
	\end{split}
	\end{equation}
\end{subequations}
where $n_{m}$ is the population of band $m$. $T_2$ is the dephasing time; $S_{mm'}(\mathbf{K},t)=\int_{-\infty}^{t}\varepsilon_{mm'}(\mathbf{K}+\mathbf{A}(t'))dt'$ is the classical action; $\varepsilon_{mm'}=E_{m}-E_{m'}$ is the band gap between bands $m$ and $m'$; $\Omega_{mm'}(\mathbf{K},t)=\mathbf{F}(t)\mathbf{d}_{mm'}(\mathbf{K}+\mathbf{A}(t))$
is the Rabi frequency where $\mathbf{d}_{mm'}(\mathbf{k})$ is the transition dipole moment. $\mathbf{K}$ is obtained from crystal momentum $\mathbf{k}$ by $\mathbf{K}=\mathbf{k}-\mathbf{A}(t)$.
\begin{subequations}
	Then the intraband current $\mathbf{j}_{ra}$ and interband current $\mathbf{j}_{er}$ can be given by
	\begin{equation}
	\mathbf{j}_{ra}(t)=\sum_{m}\int_{\overline{\mathrm{BZ}}}\mathbf{v}(\mathbf{K}+\mathbf{A}(t))n_{m}(\mathbf{K},t)d^{3}\mathbf{K},
	\end{equation}
	\begin{equation}
	\mathbf{j}_{er}(t)=\frac{d}{dt}\sum_{m\ne m'}\int_{\overline{\mathrm{BZ}}}\mathbf{p}_{mm'}(\mathbf{K},t)d^{3}\mathbf{K},
	\end{equation}
\end{subequations}
where $\mathbf{v}_{m}(\mathbf{k})=\nabla_{\mathbf{k}}E_{m}(\mathbf{k})$
is the band velocity, and the polarization $\mathbf{p}_{mm'}(\mathbf{K},t)$ is
defined as
\begin{equation}
\mathbf{p}_{mm'}(\mathbf{K},t)=\mathbf{d}_{mm'}\pi_{mm'}(\mathbf{K},t)e^{iS_{mm'}}+c.c..
\end{equation}
Then the high harmonic spectrum is obtained by the modulus square of the Fourier transform of $\mathbf{j}_{ra}$ and $\mathbf{j}_{er}$. Note that the current is multiplied by a Hann window  before the Fourier transform.

The elliptically polarized laser field $\mathbf{F}=F_{x}\mathbf{\hat{e}}_{x}+F_{y}\mathbf{\hat{e}}_{y}$ is given by
\begin{subequations}
	\begin{equation}
	F_{x}(t)=\frac{1}{\sqrt{1+\varepsilon^{2}}}F_{0}\cos^{2}(\frac{\omega_{0}t}{2n})\cos(\omega_{0}t+\phi),
	\end{equation}
	\begin{equation}
	F_{y}(t)=\frac{\varepsilon}{\sqrt{1+\varepsilon^{2}}}F_{0}\cos^{2}(\frac{\omega_{0}t}{2n})\sin(\omega_{0}t+\phi),
	\end{equation}
\end{subequations}
where $F_{0}$ is the peak of electric field inside matter, $\omega_{0}$ the frequency, $\varepsilon$ the ellipticity, and $\phi$ the carrier-envelope phase (CEP).  $n$ is the number of total cycles and is set as 20 in all of our analysis. The negative (positive) ellipticity $\varepsilon$ is defined as the left-handed (right-handed) helicity.

\subsection{Band structure of ZnO model}
\begin{figure}[t]
	\includegraphics[width=0.5\textwidth]{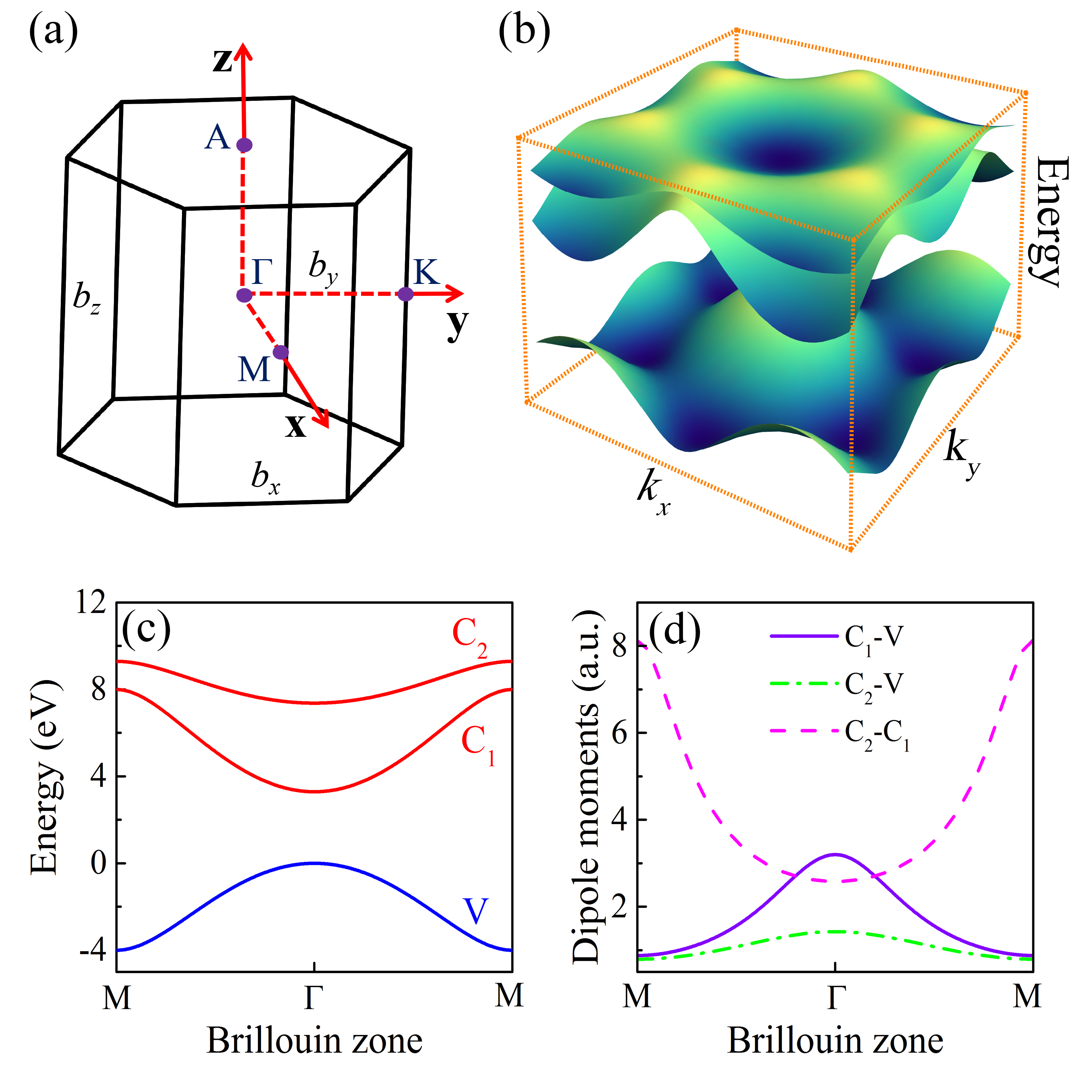}
	\caption{(Color online) (a) Brillouin zone of ZnO (wurtzite structure). (b) Band structure in the $k_{z}=0$ plane. A single valence band and two conduction bands are considered. (c) Band structure along $\Gamma-M$ direction. (d) The amplitude of the transition dipole $d^{x}_{mm'}$ between each pair of bands considered.}
	\label{fig1}
\end{figure}

The wurtzite ZnO has a hexagonal lattice \cite{UOZUGR2005,PhysRevLett.120.253201}, the first Brillouin zone is shown in Fig. \ref{fig1}(a) and the coordinates are established so that $\hat{\mathbf{x}}||\Gamma-M$, $\hat{\mathbf{y}}||\Gamma-K$, $\hat{\mathbf{z}}||\Gamma-A$ (optical axis); lattice constants $(a_{x},a_{y},a_{z})=(5.32,6.14,9.83)$ a.u. \cite{PhysRevB.91.064302,Goano2007}; reciprocal space wave vector $(b_{x},b_{y},b_{z})=(\pi/\sqrt{3}a_{x},4\pi/3\sqrt{3}a_{y},2\pi/a_{z})$. In this paper the major axis of polarization is chosen as $\Gamma-M$ direction. To investigate ellipticity dependence of HHG from ZnO, we take a single valence band (VB) and two conduction bands (CB1 and CB2)
\begin{equation}
E_{m}(\mathbf{k})=E_{m,xy}(k_{x},k_{y})+E_{m,z}(k_{z}).
\end{equation}
Here,
\begin{table}[b]
	\setlength{\tabcolsep}{6mm}{
		\centering
		\caption{Band structure parameters of ZnO for the hexagonal valence and conduction bands.}\label{table1}
		\begin{tabular}{lccc}
			\hline\hline
			&VB&CB1&CB2\\
			\hline
			$t$&2.38&-2.38&-1.00\\
			$t'$&-0.020&-0.020&-0.008\\
			$u$&27.1&27.1&27.1\\
			$p$&-7.406&10.670&10.500\\
			$q$&4.0&3.3&3.5\\
			$\alpha{_0}{^z}$&-0.0059&-0.0435&-0.0335\\
			$\alpha{_1}{^z}$&0.0059&0.0435&0.0335\\
			\hline\hline
	\end{tabular}}
\end{table}

\begin{subequations}
	\begin{equation}
	E_{m,xy}(k_{x},k_{y})=\frac{t_{m}\sqrt{f+q_{m}}+t'_{m}f+p_{m}}{u},
	\end{equation}
and
	\begin{equation}
	E_{m,z}(k_{z})=\sum_{j=0}^{1}\alpha_{m,j}^{z}\cos(jk_{z}a_{z}),
	\end{equation}
\end{subequations}
where
\begin{equation}
f=2\cos(\sqrt{3}k_{y}a_{y})+4\cos(\frac{\sqrt{3}}{2}k_{y}a_{y})\cos(\sqrt{3}k_{x}a_{x}).
\end{equation}
This analytical form of energy bands guarantees the hexagonal symmetry and size of the Brillouin zone. The band parameters ($t_m,t'_m,p_m,q_m$) listed in Table \ref{table1} are obtained from the data of nonlocal empirical pseudopotential method (NL-EPM) bands \cite{Goano2007,hawkins2016high} using least squares fitting; the nearest-neighbor expansion parameters \cite{PhysRevB.91.064302} ($\alpha_0^z,\alpha_1^z$) are used for $\Gamma-A$. The band structure of $k_{z}=0$ plane is presented in Fig. \ref{fig1}(b), and the energy bands along $\Gamma-M$ are shown in Fig. \ref{fig1}(c), where the minimum bandgap at the $\Gamma$ point is given by $\varepsilon_{g}=0.1213$ a.u. (3.3 eV).

The $\mathbf{k}$-dependent dipole element $\mathbf{d}(\mathbf{k})$ is calculated by \cite{PhysRevA.92.033845}
\begin{equation}
d_{mm'}^j(\mathbf{k})=\sqrt{\frac{E_{p,j}}{2\varepsilon^{2}_{mm'}}},
\end{equation}
where $j=x,y,z$ and $E_{p,j}$ are the Kane parameters \cite{Vurgaftman2011,UOZUGR2005,Yan2012}, we use $E_{p,x}=E_{p,y}=0.302$ a.u. and $E_{p,z}=0.375$ a.u. for ZnO model. The amplitude of transition dipole $d_{mm'}^x(\mathbf{k})$ is plotted in Fig. \ref{fig1}(d). Here the dipole phase is neglected. Note that the transition dipole and band structure obtained from different methods could result in quantitative differences of the high-harmonic spectra \cite{PhysRevA.85.043836,PhysRevA.94.013846}, for example, the even harmonics are not available \cite{PhysRevLett.120.253201,PhysRevA.96.053850}. But these differences do not change our qualitative conclusions in this paper.

\subsection{Saddle-point equations and recollisions}
In order to understand the ellipticity dependence of HHG from ZnO crystals, we use the electron-hole recollision model and solve the saddle-point equations \cite{PhysRevB.91.064302,vampa2016role}:
\begin{subequations}
	\begin{equation}
	\label{eq9a}
	\int_{t_{b}}^{t}\Delta\mathbf{v}[\mathbf{k}-\mathbf{A}(t)+\mathbf{A}(\tau)]d\tau=0,
	\end{equation}
	\begin{equation}
	\varepsilon_{mm'}[\mathbf{k}-\mathbf{A}(t)+\mathbf{A}(t_b)]-\frac{i}{T_2}=0,
	\label{eq9b}
	\end{equation}
	\begin{equation}
	\varepsilon_{mm'}(\mathbf{k})-\omega+\frac{i}{T_2}=0,
	\label{eq9c}
	\end{equation}
\end{subequations}
where $\Delta\mathbf{v}(\mathbf{k})=\nabla_\mathbf{k}\varepsilon_{mm'}(\mathbf{k})=\mathbf{v}_e-\mathbf{v}_h$. Equation (\ref{eq9a}) can be further transformed into $\Delta{\mathbf{x}_e}-\Delta{\mathbf{x}_h}=0$, which implies that high harmonics are emitted only when the electron and its associated hole recollide. In order to obtain a real solution and simplify the discussion, we set $T_{2}=\infty$ and ignore the influence of the tunnelling step. In this case, equation (\ref{eq9b}) can be solved with
\begin{equation}
\mathbf{k}=\mathbf{k}_0+\mathbf{A}(t)-\mathbf{A}(t_b),
\end{equation}
where $\mathbf{k}_0$ is the crystal momentum at minimum band gap. In our analysis, $\mathbf{k}_{0}=0$, thus $\mathbf{k}=\mathbf{A}(t)-\mathbf{A}(t_b)$. Equation (\ref{eq9c}) indicates that a harmonic photon with the energy equal to the band gap is emitted when the electron recombines with its associated hole.

In the driving of elliptically polarized field, carriers move in two-dimensional space. We relax the recollision condition as
\begin{equation}
\Delta S\equiv|\Delta\mathbf{x}_e-\Delta\mathbf{x}_h|\leq L_r,
\end{equation}
to satisfy the investigation of ellipticity dependence, where $|\Delta\mathbf{x}_e-\Delta\mathbf{x}_h|=\sqrt{(\Delta x_{e}-\Delta x_{h})^2+(\Delta y_{e}-\Delta y_{h})^2}$ and $L_r$ is the recollision distance that can be adjusted.

\section{Two-band results}
\subsection{Ellipticity dependence transition}
\begin{figure}[t]
	\includegraphics[width=0.5\textwidth]{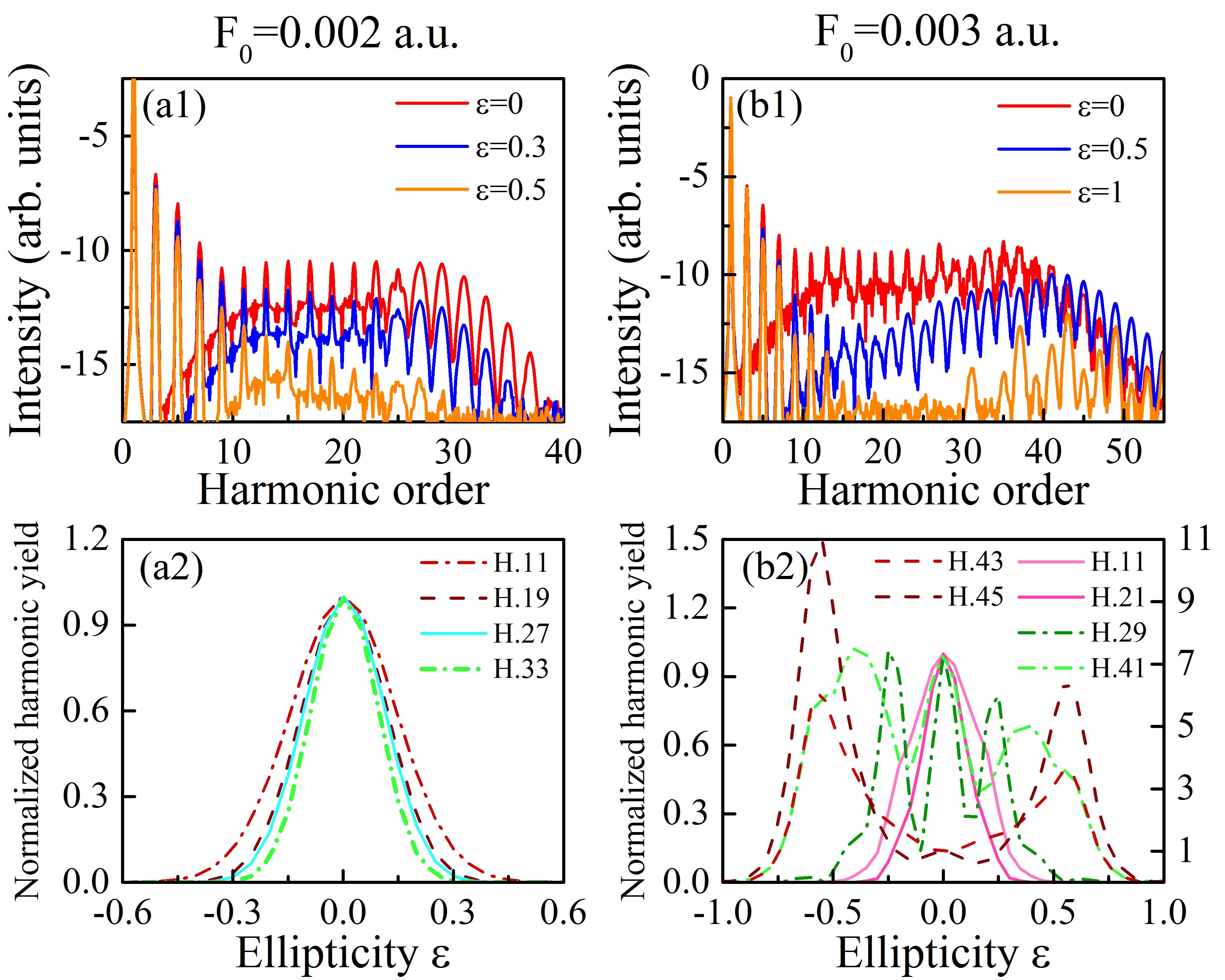}
	\caption{(Color online) Harmonic spectra and corresponding harmonic yields as a function of ellipticity for (a) $F_{0}$ = 0.002 a.u. ($I_0=1.4\times10^{11}$ W/cm$^2$) and (b) $F_{0}$ = 0.003 a.u. ($I_0=3.15\times10^{11}$ W/cm$^2$). The laser frequency $\omega$ is taken as 0.0117 a.u. (wavelength $\lambda=3.9$ $\mu$m). The dephasing time $T_2$ is set as 4 fs. In (b2), the right-hand axis is for H.43 and H.45. The asymmetric profiles of ellipticity-dependent yields are due to the neglect of transition dipole phases.}
	\label{fig2}
\end{figure}

Before solving the three-band density matrix equations we investigate the ellipticity dependence of HHG from the two-band system (VB and CB1). Figure \ref{fig2} shows the harmonic spectra and corresponding harmonic yields as a function of ellipticity for two field strengths (a) $F_0$ = 0.002 a.u. (intensity $I_0=1.4\times10^{11}$ W/cm$^2$) and (b) $F_0$ = 0.003 a.u. ($I_0=3.15\times10^{11}$ W/cm$^2$). For $F_0$ = 0.002 a.u., it is noted that both the intensity and the cutoff energy of HHG spectra [Fig. \ref{fig2}(a1)] monotonically decrease with increasing ellipticity, and at the same time, the yields of harmonics [Fig. \ref{fig2}(a2)] have a Gaussian profile. Moreover, the yields of higher-order harmonics drop more rapidly with increasing ellipticity. All these behaviors are in good agreement with the experimental observation in bulk ZnO \cite{Ghimire2010}. Increasing the field strength up to $F_0$ = 0.003 a.u., the result varies dramatically and the HHG spectra behave differently with increasing ellipticity. For example, for $\varepsilon$ = 0.5 shown in Fig. \ref{fig2}(b1) as blue line, the lower-order harmonics are suppressed dramatically, but the higher-order harmonics are enhanced obviously. This feature is more clear in Fig. \ref{fig2}(b2), which shows the normalized harmonic yields as a function of the ellipticity. While the lower-order harmonics (H.11 and H.21) follows roughly the Gaussian profile, the higher-order ones show non-monotonous behaviors, and the maximum locates at finite ellipticity, e.g., for harmonics 43 and 45, which exhibit anomalous ellipticity dependence. Without considering the details, the overall features shown in Fig. \ref{fig2}(b) are reminiscent of the experimental results (Fig. 4 of Ref. \cite{You2016}) and the simulation of time-dependent density functional theory (TDDFT) (Figs. 2 and 5 of Ref. \cite{Nicolas2017}) for single-crystal MgO in which the major polarization axis is along Mg-O direction.

The above results indicate that the harmonics can exhibit an anomalous ellipticity dependence as long as the laser field is strong enough. That is to say, one type of ellipticity dependence can be converted into another type by increasing field strength. We name this conversion as ellipticity dependence transition (EDT) which has not been observed in gaseous media. Further, we find that the EDT can also be reproduced by increasing the laser wavelength. For what was discussed above, we can draw two conclusions. First, the vector potential of incident laser plays a key role in anomalous ellipticity dependence of HHG. Second, EDT seems to bridge the gap between different types of ellipticity dependence.

\subsection{Semiclassical analysis}
\begin{figure}[t]
	\includegraphics[width=0.5\textwidth]{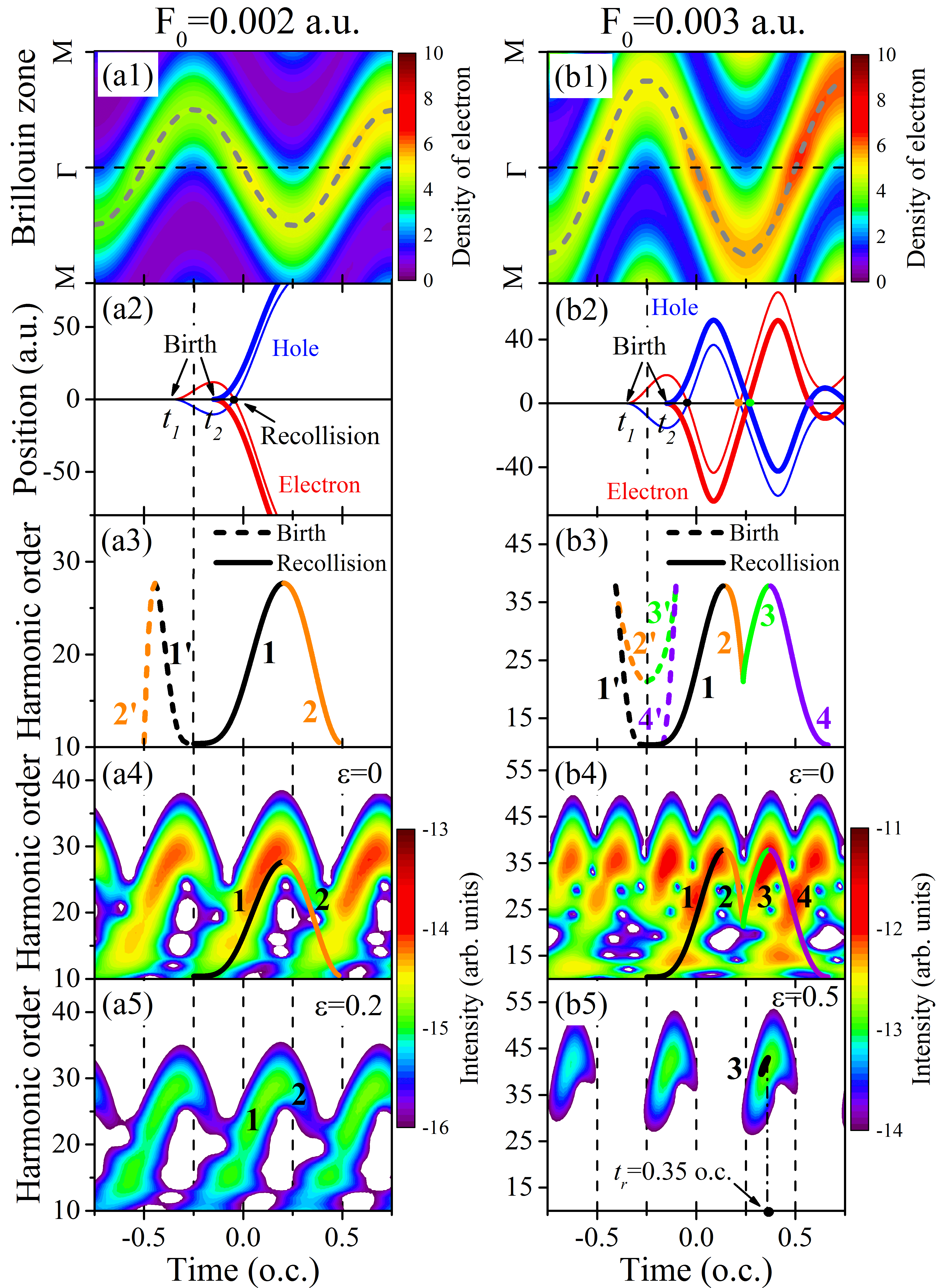}
	\caption{(Color online) Left column: $F_0 = 0.002$ a.u. ($I_0=1.4\times10^{11}$ W/cm$^2$) and Right column: $F_0 = 0.003$ a.u. ($I_0=3.15\times10^{11}$ W/cm$^2$). (a1) and (b1) Time-dependent densities of conduction electrons along $\Gamma-M$ direction. The gray dashed lines represent the vector potentials of incident lasers. (a2) and (b2) Time-dependent positions of electrons (red lines) and holes (blue lines) which are born before ($t_2=-0.15$ o.c., thick lines) and after ($t_1=-0.35$ o.c., thin lines) the peak of field. (a3) and (b3) Harmonics as a function of birth time (dashed lines) and recollision time (solid lines) calculated by the recollision model. (a4) and (b4) The time-frequency distribution of the harmonics, in which the classical trajectories are the same as those in (a3) and (b3), respectively. While all above results are calculated by taking $\varepsilon = 0$, (a5) and (b5) Same as (a4) and (b4) except the ellipticity $\varepsilon$. The color scale is logarithmic.}
	\label{fig3}
\end{figure}

The underlying physics can be understood based on time-frequency analysis and recollision model \cite{PhysRevB.91.064302,vampa2016role}. We first consider the case of linearly polarized field, namely $\varepsilon=0$, in order to confirm our calculation method. For the relative weak field strength $F_0$ = 0.002 a.u., the excited carriers are unable to go beyond the boundaries of the Brillouin zone as shown in Fig. \ref{fig3}(a1). In this case, only the pairs of electron and hole born after the peak of the field can recombine each other as seen in Fig. \ref{fig3}(a2), similar to the recollision model of the atomic case. In Fig. \ref{fig3}(a3), the classical trajectories calculated by recollision model are presented, where trajectory 1 (2) denotes \emph{short} (\emph{long}) trajectory, respectively. The time-frequency result shown in Fig. \ref{fig3}(a4) is in good agreement with classical result, which confirms the validity of the classical recollision mechanism. The quantum paths are numbered as the same as the corresponding classical trajectories.

Increasing the field strength up to $F_0$ = 0.003 a.u., an important change is that the excited pairs of electron and hole can travel across the boundaries of Brillouin zone (point M) as shown in Fig. \ref{fig3}(b1), which makes the DBO possible. As a result, the multiple oscillation of pairs in real space \cite{Foldi2013} happens, which increases dramatically the recollision possibility, regardless of the pairs born before (thick lines) or after (thin lines) the peak of field [see Fig. \ref{fig3}(b2)]. The classical trajectories shown in Fig. \ref{fig3}(b3) look more complicated, but it is clear that the trajectories 3 and 4 are new ones, which appear due to the recollisions of the pairs born before the peak of field with the help of DBO. Likewise, the time-frequency result shown in Fig. \ref{fig3}(b4) further validates the classical recollision model. In addition, due to the interference of these multiple quantum paths the multiple-peak structure of harmonics (red line) shown in Fig. \ref{fig2}(b1) is also understandable \cite{PhysRevLett.100.143902}.

In the following, we turn to the laser fields with finite ellipticity, as shown in Fig. \ref{fig3}(a5) and \ref{fig3}(b5). For $F_0$ = 0.002 a.u., the time-frequency distribution has no apparent change except for suppressed intensity, see Fig. \ref{fig3}(a5). However, dramatic change is observed for $F_0$ = 0.003 a.u. as $\varepsilon$ = 0.5 is taken. While the paths 1 and 2 vanish completely, however, see \ref{fig3}(b5), path 3 and part of path 4 remain survived although their intensities become weakened. This implies that quantum paths born before the peak of field (paths 3 and 4) are less sensitive to ellipticity than those born after the peak (paths 1 and 2). Taking the recollision distance as 10 a.u., the 2D recollision model gives the similar result that only part of trajectory 3 [short black line in Fig. \ref{fig3}(b5)] survived. Moreover, it is very obvious that path 3 moves towards higher frequency and its cutoff frequency even extends to the 43rd harmonic. One notes that the paths 3 and 4 appear under the help of DBOs as pointed out above, which shows an important role played by the DBO in the ellipticity-dependent harmonics.

After having clarified the classical trajectories and obtained the time-frequency behaviors for different field strengths, let us come back to the anomalous ellipticity dependence presented in Fig. \ref{fig2}(b2). Since paths 2 and 4 contribute weakly to the harmonics, we focus on paths 1 and 3 below. From Fig. \ref{fig3} (b4), it is seen that the harmonics below order 21 are originated only from the path 1, which is suppressed strongly by the finite ellipticity. As a consequence, the yields of harmonics 11 and 21 in Fig. \ref{fig2}(b2) exhibit Gaussian profile as a function of ellipticity. Harmonics between order 21 and 37 (cutoff) contain the contributions both from paths 1 and 3. However, path 3 is less sensitive to ellipticity and moves towards higher frequency at finite ellipticity, this is why that the harmonic 29 shows a three-peak structure. After order 37, harmonics exceed the cutoff position predicted by the classical model in the driving of linearly polarized field, and the ellipticity-dependent yields of harmonics evolve from a rough three-peak structure (see harmonic 41) into a two-peak structure (see harmonics 43 and 45). This behavior can be due to the competition between two opposite effects acted on the path 3 as ellipticity increases: one is (i) the cutoff extension, and the other is (ii) the overall suppression of quantum paths. While the consequence of (i) enhances intensities of harmonics after the cutoff, and the effect (ii) weakens the intensities of all harmonics. The interaction of these two factors can lead to the two-peak structure of the harmonics 43 and 45, which show maximum at a finite ellipticity of $\varepsilon \approx 0.55$.

Have uncovered that the path 3 plays an important role in obtaining the anomalous ellipticity dependence of HHG, two questions have not clearly clarified yet, namely, (i) why the path 1 is sensitive to ellipticity but the path 3 is not and (ii) why the cutoff of the path 3 is extended when increasing ellipticity. In the following we answer these two questions.

\begin{figure}[t]
	\includegraphics[width=0.5\textwidth]{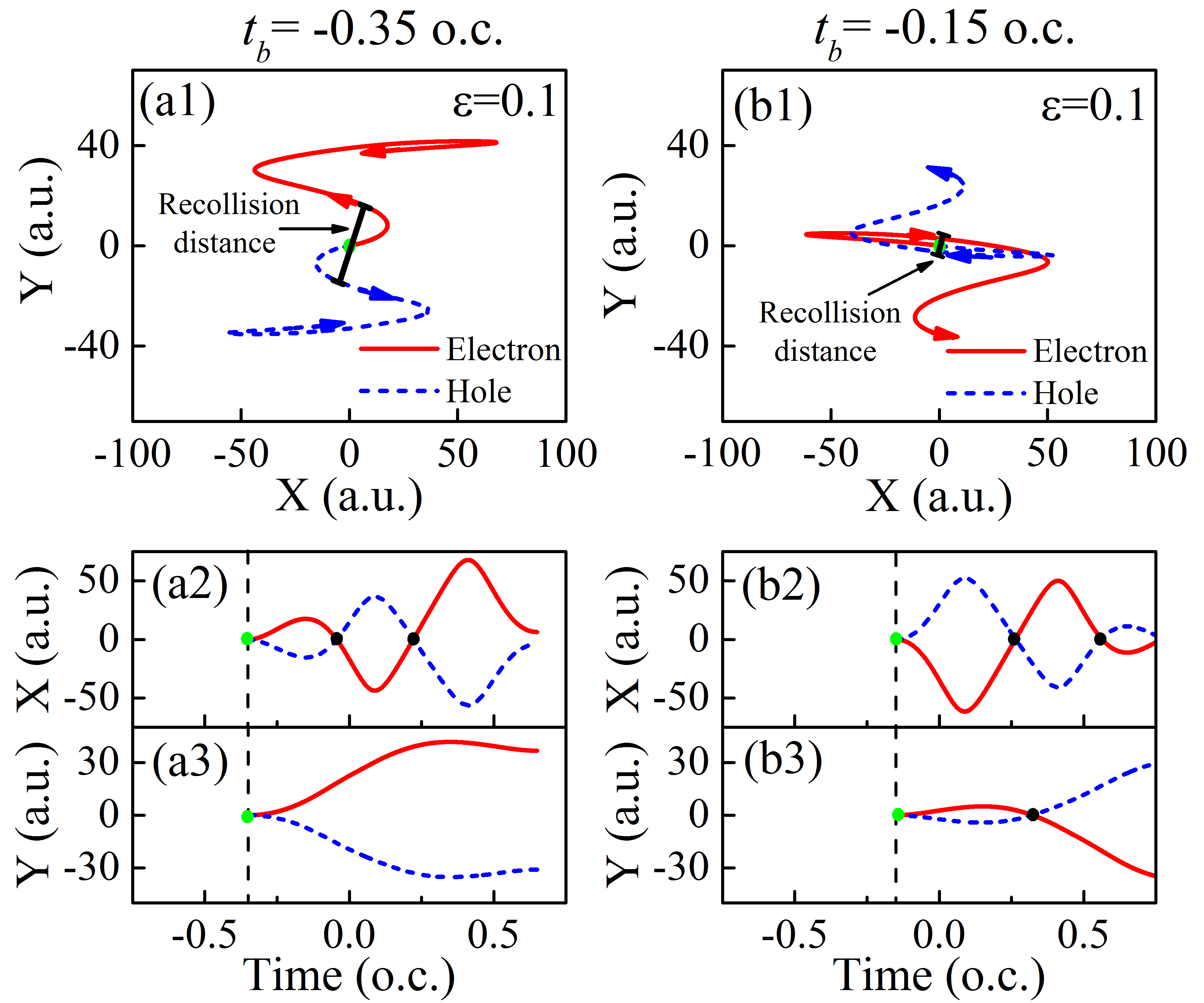}
	\caption{(Color online) (a1) and (b1) The real-space trajectories of the electron-hole pair. (a2) and (b2) The time-dependent positions of the pair in X ($\Gamma-M$) direction. (a3) and (b3) The time-dependent positions of the pair in Y ($\Gamma-K$) direction. We take the field strength $F_0$ = 0.003 a.u. and ellipticity $\varepsilon$ = 0.1. Left column: birth time $t_b$ = $-0.35$ o.c. and Right column: birth time $t_b$ = $-0.15$ o.c.. }
	\label{fig4}
\end{figure}

In order to answer the question (i), let us check carefully Figs. \ref{fig3}(b4) and \ref{fig3}(b5) again. It is found that the sensitivity of quantum paths to ellipticity is related to the birth time of the corresponding pairs. Thus we analyze real-space trajectories of pairs born after (birth time $t_b=-0.35$ o.c.) and before ($t_b=-0.15$ o.c.) the peak of field and show the result in Fig. \ref{fig4}. It is obvious that the recollision distance of pairs born at $-0.15$ o.c. is much smaller than that born at $-0.35$ o.c. as driving in the same ellipticity of laser. This can be understood as follows. For elliptically polarized field, there is a phase difference of $\pi/2$ between the electric vector in X- and Y-direction. Thus the electric field before the peak of $F_{x}$ is synchronous with the field after the peak of $F_{y}$ and vice versa. Consequently, the 2D trajectories of pairs can be seen as a simple synthesis of time-dependent positions in parallel (X) and perpendicular (Y) direction. For $F_{0}=0.003$ a.u. and $\varepsilon=0.1$, we know that $F_{x,0}\approx 0.003$ a.u., $F_{y,0}=\varepsilon F_{x,0}\approx 0.0003$ a.u.. In X-direction, the field strength is strong enough to generate DBO, thus electron and hole in X-direction can encounter with each other multiple times during one optical cycle [Figs. \ref{fig4}(a2) and \ref{fig4}(b2)], regardless of the pairs born before or after the peak of field. However, the field strength in Y-direction is too weak to result in DBO so that the movement of pairs in this direction is similar to the atomic case. When pairs are excited after the peak of $F_x$ (before the peak of $F_y$), as shown in Fig. \ref{fig4}(a3), the distance between electron and hole in Y-direction becomes larger and larger, no matter how they oscillate in X-direction. This leads to that the path 1 is sensitive to ellipticity. For the pairs born before the peak of $F_x$ (after the peak of $F_y$), see Figs. \ref{fig4}(b2) and \ref{fig4}(b3), the excited electron can reencounter with its associated hole both in X- and Y-direction. It leads to that the electron and hole born before the peak of $F_x$ are hard to be pulled apart. This is the reason that the path 3 is less sensitive to ellipticity.

\begin{figure}[t]
	\includegraphics[width=0.5\textwidth]{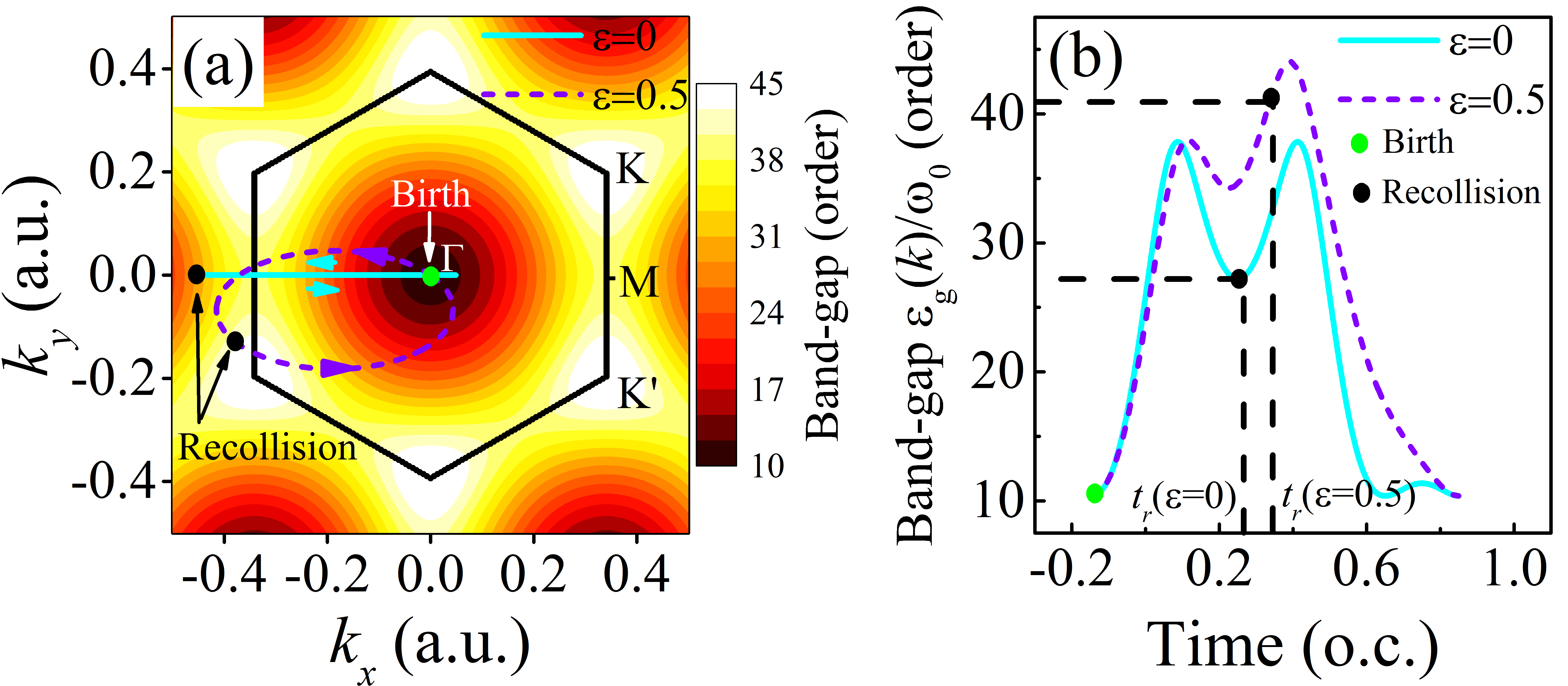}
	\caption{(Color online) (a) The band-gap ($E_{\mathrm{CB1}}-E_{\mathrm{VB}}$) distribution and trajectories of electron-hole pairs born at $-0.15$ o.c. in Brillouin zone. The color scale is linear. (b) The time-dependent band-gap of pairs driving by linearly ($\varepsilon=0$) and elliptically ($\varepsilon=0.5$) polarized field.}
	\label{fig5}
\end{figure}

Now we answer the question (ii). We check the pair born at $t_b=-0.15$ o.c.. Figure \ref{fig5}(a) shows in $\mathbf{k}$-space the band-gap distribution and trajectories of the electron-hole pair for linearly ($\varepsilon=0$) and elliptically ($\varepsilon=0.5$) polarization. In Fig. \ref{fig5}(b), we show the band-gap size experienced by the pair as a function of time. One notes that with increasing ellipticity the electron-hole pair goes across the region with larger band-gap due to the anisotropic ZnO band structure (the band-gap at point K is larger than at point M). Moreover, the recollision time $t_r$ of the electron-hole pair is delayed from 0.27 o.c. ($\varepsilon=0$) to 0.35 o.c. ($\varepsilon=0.5$) when the ellipticity increases from 0 (dashed line) to 0.5 (solid line). These two factors clearly lead to that the harmonic emitted by the electron-hole pair at the birth time of $t_b=-0.15$ o.c. elevates the order from 27 to 41 when ellipticity is changed from 0 to 0.5. As a result, the path 3 moves toward the higher frequency as increasing ellipticity. This is the reason that the cutoff of the path 3 becomes larger at finite ellipticity.

\section{Three-band results}

\begin{figure}[b]
	\includegraphics[width=0.5\textwidth]{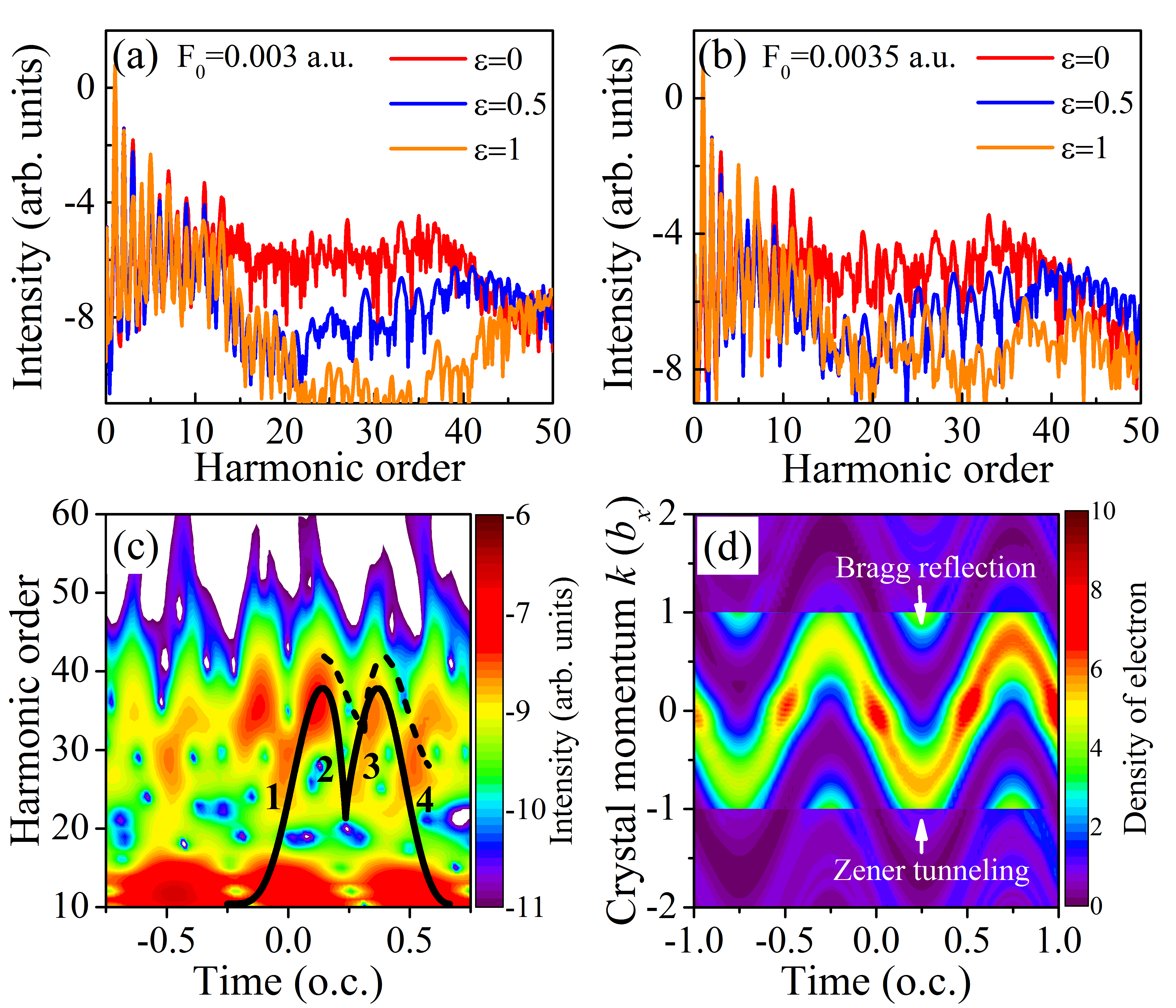}
	\caption{(Color online) Harmonic spectra as a function of ellipticity for the three-band ZnO model. (a) $F_{0}$ = 0.003 a.u. ($I_0=3.15\times10^{11}$ W/cm$^2$) and (b) $F_{0}$ = 0.0035 a.u. ($I_0=4.29\times10^{11}$ W/cm$^2$). The laser frequency $\omega$ is taken as 0.0117 a.u. (wavelength $\lambda=3.9$ $\mu$m). The dephasing time $T_2$ is set as 8 fs. (c) Time-frequency distribution of harmonics. The color scale is logarithmic. The black lines are classical trajectories calculated by three-band recollision model of ZnO. The solid lines (dashed lines) correspond to the recollision between CB1 (CB2) and VB. (d) Time-momentum distribution of the population of the two CBs (CB1 and CB2). The color scale is linear. In (c) and (d), laser parameters are the same as the linearly polarized field in (a).}
	\label{fig6}
\end{figure}

In this section, we investigate the ellipticity dependence of HHG including the second conduction band (CB2) of ZnO. For the linearly polarized laser, the energy cutoff of harmonic spectrum calculated by the three-band model [red line of Fig. \ref{fig6}(a)] is the same as that of two-band case [red line of Fig. \ref{fig2}(b1)] since the small transition dipole moments between CB2 and the VB. However, the yields of low-order harmonics (H.5 to 15) are much higher than those in Fig. \ref{fig2} owing to the strong interband polarization between CB2 and CB1. The corresponding time-frequency distribution and classical trajectories are shown in Fig. \ref{fig6}(c). When the ellipticity $\varepsilon$ increases to 0.5, the anomalous ellipticity dependence still exists, see the blue line of Fig. \ref{fig6}(a). The yields of lower-order harmonics decay faster than higher-orders, but the extend of energy cutoff at a finite ellipticity cannot be observed. This is due to the electronic wavepacket splits into two parts at BZ border, as shown in Fig. \ref{fig6}(d). The small part of wavepacket transfers to a higher band, so the intensity of path 3 is slightly weaker than that of path 1 and the cutoff extension could not be observed. The large fraction reenters the BZ through Bragg reflection on the same band, so the intensity of quantum path 3 is still strong and we can also see the  anomalous ellipticity dependence. Furthermore, when the electric field is increased to $F_0=0.0035$ a.u., as shown in Fig. \ref{fig6}(b), the extend of energy cutoff for a finite ellipticity reappears. Thus the DBO on CB1 can still cause an anomalous ellipticity dependence of HHG in the three-band ZnO system.

However, for some multi-band systems \cite{PhysRevA.91.043839,PhysRevA.91.013405,hawkins2016high,JBLi2017}, the probability of Zener tunneling at the boundary of  Brillouin zone is very large, even close to 1. At this time, the intensity of the quantum path induced by DBO is quite weak, which is not sufficient to cause the anomalous ellipticity-dependent behavior of harmonics. Therefore, the probability of tunneling between different bands at the BZ border should be carefully checked if one wants to observe the EDT in other realistic systems.

\section{Summary}
We investigate the ellipticity dependence of HHG from ZnO cystal by solving the density matrix equations. It is found that the dynamical Bloch oscillation, which can be excited by a strong vector potential of laser, plays an important role in generating an anomalous ellipticity-dependent behavior of HHG. The availability of the DBO induces new quantum paths, which, on the one hand, are less sensitive to ellipticity and on the other hand go across the region with larger band-gap due to the anisotropic band structure we study. The result not only reveals the underlying physics of ellipticity dependence transition, but to some extent provide an intuitive physical picture for understanding the anomalous ellipticity dependence observed in the single-crystal MgO. Our results could be tested in the present experimental setups, for example, by increasing the laser wavelength in bulk ZnO (as increasing field strength may lead to the material damage \cite{Ghimire2010}). 

\section*{Acknowledgements}
This work was supported by NSFC (Grant No. 11834005, No. 11874030, No. 11674139) and PCSIRT (Grant No. IRT-16R35).

\end{document}